\documentclass{IEEEtran}
\usepackage{lineno,hyperref,graphicx,multirow,amssymb,subcaption,amsmath,amsfonts}
\usepackage{algorithmic}
\usepackage{graphicx}
\usepackage{algorithm}

\begin{document}

\title{Study of Robust Adaptive Beamforming Based on Low-Complexity DFT Spatial Sampling}

\author{Saeed  Mohammadzadeh, Vitor H.Nascimento,
Rodrigo C. de Lamare, and Osman Kukrer
\thanks{Saeed Mohammadzadeh and Vitor H.Nascimento are with the Department
of Electronic Systems Engineering, University of Sao Paulo, Sao
Paulo 05508-900, Brazil, e-mails: saeed.mzadeh@ieee.org,
vitor@ieee.org, Rodrigo C. de Lamare is with CETUC, PUC-Rio
22451-900, Brazil and with the Department of Electronic Engineering,
University of York, UK, e-mail: delamare@cetuc.puc-rio.br and Osman
Kukrer is with the Department of Electrical and Electronics
Engineering, Eastern Mediterranean University, Famagusta 99450,
Turkey e-mail: osman.kukrer@emu.edu.tr.}}

\maketitle

\begin{abstract}
In this paper, a novel and robust algorithm is proposed for adaptive
beamforming based on the idea of reconstructing the autocorrelation
sequence (ACS) of a random process from a set of measured data. This
is obtained from the first column and the first row of the sample
covariance matrix (SCM) after averaging along its diagonals. Then,
the power spectrum of the correlation sequence is estimated using
the discrete Fourier transform (DFT). The DFT coefficients
corresponding to the angles within the noise-plus-interference
region are used to reconstruct the noise-plus-interference
covariance matrix (NPICM), while the desired signal covariance
matrix (DSCM) is estimated by identifying and removing the
noise-plus-interference component from the SCM. In particular, the
spatial power spectrum of the estimated received signal is utilized
to compute the correlation sequence corresponding to the
noise-plus-interference in which the dominant DFT coefficient of the
noise-plus-interference is captured. A key advantage of the proposed
adaptive beamforming is that only little prior information is
required. Specifically, an imprecise knowledge of the array geometry
and of the angular sectors in which the interferences are located is
needed. Simulation results demonstrate that compared with previous
reconstruction-based beamformers, the proposed approach can achieve
better overall performance in the case of multiple mismatches over a
very large range of input signal-to-noise ratios.
\end{abstract}


\section{Introduction}
\label{sec:introduction}

In order to enhance the desired signal arriving from the target
direction while suppressing interfering signals from other
directions, adaptive beamforming techniques have been widely applied
in radar, sonar, seismology, radio astronomy, medical imaging,
wireless communications, and other fields \cite{van2004detection}.
Conventional adaptive beamformers assume accurate knowledge of the
antenna array, of the actual array manifold, and that there is no
desired signal component in the training sample used to estimate the
noise-plus-interference covariance matrix (NPICM). However, in
practical applications, these ideal assumptions are almost
impossible to satisfy. Adaptive arrays are highly sensitive to
various mismatches, including antenna array calibration errors,
incoherent local scattering, wavefront distortion and
direction-of-arrival (DoA) error. Any such model mismatch will cause
a conventional adaptive beamformer to suffer severe performance
degradation. Moreover, the presence of the desired signal in the
training snapshots will result in significant self-nulling of the
desired signal, causing the output signal-to-interference-plus-noise
ratio (SINR) to decrease \cite{rahmani2014robust},
\cite{liu2018coprime}.

Over the past decades, a large number of robust adaptive beamforming \cite{vorobyov2003robust,rdrcb,jidf,jio,jiols,sjidf,l1stap,rrdstap,jiolcmv,rrstap,wljio,okspme,mcg,eljio,wcccm,ccmjio,ccmavf,arh,kaesprit,rdrls,lrcc} approaches have been proposed, which can mitigate the effects of model mismatches and improve the robustness of beamformers. In \cite{mestre2006finite,kukrer2014generalised}, the diagonal loading method was used to reduce the sensitivity to the desired signal. The main shortcoming of this approach is that it is not clear how to determine optimal values of the diagonal loading level for different scenarios. In \cite{huang2012modified}, eigen-subspace decomposition and projection techniques were used to improve the robustness of adaptive beamforming at high signal-to-noise ratios (SNRs). However, these methods suffer from serious performance degradation at low SNRs, where the desired signal subspace may be contaminated by the noise subspace, and they have high computational complexity, especially for a large array \cite{xie2014fast}. In \cite{vorobyov2003robust} and \cite{yu2010robust}, a worst-case-based technique was proposed to achieve good performance. However, with this approach, it is very difficult to obtain the steering vector (SV) mismatch and the error bound in practical applications. Moreover, at high SNRs, the performance of this method will severely degrade in the presence of array SV errors. The uncertainty set-based algorithms estimate the desired signal SV based on the elliptical and spherical uncertainty set of the (signal-of-interest) SOI steering vector by solving an optimization problem \cite{nai2011iterative}. However, their performance is mainly determined by the uncertain parameter set and it is difficult to select the optimal factor in practice \cite{yang2017modified}. In addition, the design of adaptive beamforming techniques based on these principles has some drawbacks such as their ad hoc nature, high probability of subspace swap at low SNR, presence of the SOI component in the sample covariance matrix (SCM) and high computational complexity.

In the last decade, a new approach has been put forward in which the impact of the SOI component is removed from the SCM by reconstructing the NPICM. The NPICM in \cite{gu2012robust} is reconstructed based on the Capon spectral estimator by integrating over an angular sector that excludes the DoA of the SOI, while the desired signal SV is estimated by solving a quadratically constrained quadratic programming (QCQP) problem. This method shows reasonable performance, but is sensitive to large DoA mismatches \cite{yuan2017robust,mohammadzadeh2018modified}. Several NPICM-based beamformers were then proposed, such as low-complexity methods that reconstruct the NPICM by subtracting the reconstructed DSCM from the SCM \cite{ruan2014robust}, and the sparse reconstruction method \cite{gu2014robust}.

In \cite{huang2015robust}, an annular uncertainty set was used to constrain the SVs of the interferences during NPICM reconstruction. The performance of this approach is very close to that of the beamformer in \cite{gu2012robust}. However, because the NPICM is reconstructed by integrating over a complex annular uncertainty set, this approach has a great disadvantage in computational complexity. Several weighted subspace-fitting-based NPICM reconstruction beamformers were proposed \cite{chen2018robust}, which are especially designed to mitigate the effect of sensor position errors by compensating for the estimated sensor position errors in the NPICM reconstruction.

{The beamformer in \cite{chen2015robust} utilises the same approach in \cite{gu2012robust} to reconstruct the NPICM while it uses correlations between the presumed steering vector of the SOI and the eigenvectors of the sample covariance matrix to estimate the desired signal SV. Thus, this approach can not eliminate the subspace swap error in the case of low SNRs.} In \cite{zhang2016interference} a robust beamforming algorithm has been developed based on the IPNC matrix reconstruction using spatial power spectrum sampling (SPSS). This method has lower computational complexity, but its performance is degraded as the number of sensors is decreased. The algorithm in \cite{mohammadzadeh2018adaptive} jointly estimates the theoretical IPNC matrix using the eigenvalue decomposition of the received signal covariance matrix and the mismatched steering vector using the output power of the beamformer. In \cite{zheng2018covariance} a procedure analogous to those of \cite{gu2012robust} and \cite{li2003robust} is used to reconstruct the IPNC matrix and the desired signal steering vector estimation. However, the accuracy of the interference steering vector estimation is related to an ad hoc parameter.
The results of \cite{gu2012robust} demonstrate that the resulting Capon beamformer allows for good performance in the case of SOI array steering vector errors. However, the analysis did not account for typically present interference array steering vector errors or arbitrary SOI array steering vector mismatches \cite{somasundaram2014degradation}. Besides, the accuracy of the Capon spatial spectrum degrades severely when coherent signals (with line spectra) exist \cite{wang2016robust}. In order to avoid this problem, a very recent algorithm in \cite{mohammadzadeh2020maximum} based on the NPICM and DSCM was proposed which estimates all interference powers as well as  the  desired signal power using the principle  of maximum entropy power spectrum with low computational complexity. In \cite{sun2021robust}, an new algorithm is proposed based on the steering vector estimation utilizing gradient vector which is orthogonal to assumed steering vector. Then, the IPNC matrix is reconstructed by estimated interference steering vectors and corresponding powers. Although this method is robust against some mismatches, it computational complexity is high. {The beamformer in \cite{zhu2020robust} utilizes subspace orthogonality to reconstruct the NPICM. In \cite{zhang2020rcb}, a design of SOI power estimator to formulate the SV optimization problem with an uncertainty set constraint is proposed which is different from the  conventional Capon estimator in which the desired signal SV is optimized with the Capon power estimator.}

Motivated by the above fact, in this paper, different from the previous NPICM reconstruction methods, we first develop a method based on the idea of reconstructing the autocorrelation sequence (ACS) of a random process from a set of measured data, and then taking the Discrete Fourier Transform (DFT) to obtain an estimate of the power spectrum, which is denoted reconstruction based on the DFT (REC-DFT). The reconstructed sequence is obtained from the first column and the first row of the SCM after averaging all of its diagonals. The DFT coefficients corresponding to the angles within the noise-plus-interference region are used to reconstruct the NPICM, while the DSCM is estimated by identifying and removing the noise-plus-interference component from the SCM.\\
The paper's contributions can be summed up as follows:
{\begin{enumerate}
\item[1.] A low computational complexity robust adaptive beamforming, called REC-DFT, based on the autocorrelation sequence of a random process is developed.
\item[2.] The NPICM is reconstructed directly and without the need to estimate the power of the interferences and the corresponding array steering vectors.
\item[3.] The DSCM is reconstructed based on the elimination of the noise-plus-interference components from SCM and the desired signal SV is estimated by multiplication of the reconstructed DSCM an the assumed SV.
\item[4.] We demonstrate that the suppression of interference is significantly enhanced by the proposed REC-DFT beamforming method compared to the existing methods in the literature.
\end{enumerate}}
\section{The Signal Model and Background}\label{sec2}
Consider a uniform linear array (ULA) consisting of $ N $ sensors that receive $ L+1 $ narrowband far-field sources. The $ {N \times1} $ complex array observation data vector at the $k$th snapshot can be expressed as
\begin{align}\label{x}
\mathbf{{x}}(k)=\mathbf{{x}}_s(k)+\mathbf{{x}}_l(k)+\mathbf{{n}}(k),
\end{align}
where $\mathbf{{x}}_s(k)=s_0(k)\mathbf{{a}}_s$, $\mathbf{x}_l(k)=\sum_{l=1}^{L}s_l(k)\mathbf{{a}}_l$ are the components of the desired signal and the interferences. The additive Gaussian noise vector $\mathbf{{n}}(k)$ is spatially independent from the interferences and desired signal. Also, $s_0(k)$ is the waveform of the desired signal, and $ \mathbf{{a}}_s $ is the actual SV of the SOI. Furthermore, $ \mathbf{{a}}_l $ is the interference signal SV and $s_l(k)$ is the corresponding waveform at the $k$th snapshot. The array SV corresponding to the direction of the signals is defined as
\begin{align}
    \mathbf{a}(\theta)=\Big[1, e^{-j \theta}, \cdots, e^{-j  (N-1)\theta} \Big]^T,
\end{align}
where $\theta=\pi \sin \phi$ (assuming half-wavelength sensor spacing), $\phi$ is the angle of arrival and $(\cdot)^T$ denotes transposition. For the rest of paper, it is assumed that, $\mathbf{a}(\theta_l)=\mathbf{a}_l$ and  $\mathbf{a}(\theta_s)=\mathbf{a}_s$ where $\theta_l$ and $\theta_s$ denote the DoA of the interference and SOI signal, respectively.\\
To quantitatively measure the interference suppression capability of adaptive beamformers, the output SINR is defined as
\begin{align}\label{SINR}
\mathrm{SINR} \triangleq \dfrac{\sigma^{2}_s |\mathbf{{w}}^H \mathbf{{a}}_s|^2 }{\mathbf{{w}}^H \mathbf{{R}}_{\text{ipn}}\mathbf{{w}}},
\end{align}
where $ \mathbf{{w}}=[w_{1}, \ \cdots, \ w_{N} ]^{T} $, $ (\cdot)^{H} $ and $ \sigma^{2}_s=E\{ |s_0(k)|^2 \} $ are respectively the beamformer weight vector, the Hermitian transpose operator and the desired signal power while $E\{ \cdot \}$ stands for the statistical expectation operator. Also, $ \mathbf{{R}}_{\text{ipn}} $ is the NPICM which is given by
\begin{align}\label{theoritical Rin}
\mathbf{{R}}_{\text{ipn}} & \triangleq \mathbf{E} \big\{ (\mathbf{{x}}_l(k)+\mathbf{n}(k))(\mathbf{{x}}_l(k)+\mathbf{n}(k))^H \big \} \nonumber \\&=\mathbf{{R}}_i+\mathbf{{R}}_n= \sum_{l=1}^{L}\sigma_l^2\mathbf{{a}}_l\mathbf{{a}}_l^H+\sigma_n^2 \mathbf{{I}},
\end{align}
where $ \sigma_l^2 $ $ \ (l=1,...,L) $ is \textit{l}{th} interference power and $ \sigma_n^2 $ denotes the noise power, and $\mathbf{{I}}$ is an identity matrix of order N. \\
The standard beamformer intends to maintain the SOI without any distortion while the noise-plus-interference components are suppressed, thereby the output SINR is maximized. The beamformer can be expressed as
\begin{align}\label{MVDR}
\underset{{\mathbf{{w}}}}{\operatorname{min}}\ \mathbf{{w}}^H \mathbf{{R}}_{\text{ipn}}\mathbf{{w}}\ \hspace{.4cm} \mathrm{s.t.} \hspace{.4cm} \mathbf{{w}}^H \mathbf{{a}}_s=1
\end{align}
The optimal solution to (\ref{MVDR}) are the weights of the standard Capon beamformer
\begin{align}\label{optimal wegight vector}
\mathbf{{w}}_{\mathrm{opt}}=\dfrac{\mathbf{{R}}_{\text{ipn}}^{-1} \mathbf{{a}}_s}{\mathbf{{a}}_s^H \mathbf{{R}}_{\text{ipn}}^{-1}\mathbf{{a}}_s}.
\end{align}
In practical applications, the actual SV of the SOI, $ \mathbf{{a}}_s $, and the actual NPICM, $ \mathbf{{R}}_{\text{ipn}}$ are unavailable. Therefore, $ \mathbf{{a}}_s $ is usually replaced by the assumed SV, $ \bar{\mathbf{a}}_s$, and $ \mathbf{{R}}_{\text{ipn}} $ can be replaced by the SCM \cite{van2004detection}
\begin{align}\label{SCM}
\hat{\mathbf{{R}}}=\dfrac{1}{K}\sum_{k=1}^{K} \mathbf{{x}}(k)\mathbf{{x}}^H(k),
\end{align}
The total number of snapshots is $ {K} $. Also, it is well known that, the SCM converges to the theoretical covariance matrix, when $K\to \infty $ is reached, as
\begin{align}\label{Theoretical R}
    \mathbf{{R}}=\mathbf{{R}}_s+\mathbf{{R}}_{\text{ipn}}=\sigma_s^2\mathbf{{a}}_s\mathbf{{a}}_s^H+\sum_{l=1}^{L}\sigma_l^2\mathbf{{a}}_l\mathbf{{a}}_l^H+\sigma_n^2 \mathbf{{I}},
\end{align}
where $\mathbf{{R}}_s=\sigma_s^2\mathbf{{a}}_s\mathbf{{a}}_s^H$ is the DSCM.
\section{NPICM Reconstruction}\label{sec3}
In this section, we first describe analytically the NPICM reconstruction technique using the Capon spectrum estimation. Then, we develop a new robust adaptive approach (REC-DFT) that achieves near-optimal performance by addressing both the inaccurate covariance matrix problems as well as  the SV mismatches.
\subsection{Capon Based Matrix Reconstruction}
It is clear that the most important issue with NPICM reconstruction is the accuracy of the power spectrum estimate. Inaccuracies in the power spectral estimate result in distorted angular positions of interference signals, as well as their powers, which eventually lead to their insufficient suppression. The use of the Capon spectrum estimation method to reconstruct the NPICM was introduced in \cite{gu2012robust}.
Despite the many papers that built on NPICM reconstruction using the Capon estimator \cite{capon1969high}, it has never been demonstrated that the reconstruction based on the Capon estimator could be approximately equal to the NPICM. Although such a result is expected, it is not tractable to find an analytical proof, even if  the theoretical $\mathbf{R}_{\text{ipn}}$ is replaced with the estimated $ \hat{\mathbf{R}}_{\text{ipn}} $. To emphasize this, we present a somewhat heuristic argument demonstrating that the reconstruction methodology is sensible. To reconstruct the NPICM, we assume that there is one interference signal with SV $\mathbf{a}_l$, and that the SOI signal does not exist in the training data. Hence, for this case, we can rewrite the theoretical covariance matrix as follows
\begin{align}\label{App R}
\mathbf{R}_{(c)}=\sigma^2_n\mathbf{I}_\mathrm{N}+\sigma^2_l\mathbf{a}_l\mathbf{a}_l^H.
\end{align}
Moreover, there is a proof that the optimal weight vector does not change the optimal output SINR, even if the NPICM is replaced by the theoretical, $\mathbf{R}_{(c)}$ \cite{van2004detection}. Based on the Capon estimator, authors in \cite{gu2012robust} proposed an algorithm to reconstruct the NPICM as follows
\begin{align}\label{Ripn}
\mathbf{R}_{(c)\text{ipn}}=&\int_{\Theta_{\text{ipn}}} \rho(\theta) \mathbf{a}(\theta)\mathbf{a}^H(\theta) d\theta \nonumber \\ =&\int_{\Theta_{\text{ipn}}}\frac{\mathbf{a}(\theta)\mathbf{a}^H(\theta)}{\mathbf{a}^H(\theta)\mathbf{R}^{-1}_{(c)}\mathbf{a}(\theta)}d\theta,
\end{align}
where $\rho(\theta)$ is the power spectrum in the noise-plus-interference spatial domain and $\Theta_{\text{ipn}} \cup \Theta_s=[-\pi,\pi]$. It is assumed that the angular sector of the interferences, $\Theta_{\text{ipn}}$ and the location of the desired signal region, $\Theta_s$ are distinguishable. Also, $\Theta_{\text{ipn}}$ is approximated by a summation by sampling over $\Theta_{\text{ipn}}$ with step $\Delta\theta$  as
\begin{align}\label{App Ripn}
\mathbf{R}_{(c)\text{ipn}}=\sum_{p=1}^{P}\frac{\mathbf{a}(\theta_p)\mathbf{a}^H(\theta_p)}{\mathbf{a}^H(\theta_p){\mathbf{R}}^{-1}_{(c)}\mathbf{a}(\theta_p)}\Delta\theta.
\end{align}
We can express the inverse of the covariance matrix in (\ref{App R}) using the application of the matrix inversion lemma (Woodbury) as shown below
\begin{align}\label{App R invers}
\mathbf{R}^{-1}_{(c)}=\frac{1}{\sigma^2_n} \Big(\mathbf{I}_N-\frac{\mathbf{a}_l\mathbf{a}_l^H}{\gamma+\|\mathbf{a}_l\|^2} \Big),
\end{align}
where $\| \mathbf{a}_l \|^2=N$ and $ \gamma=\sigma^2_n/\sigma^2_l $. The denominator of (\ref{App Ripn}) is re-written by substituting of (\ref{App R invers}) as follows
\begin{align}\label{R4}
\mathbf{a}^\mathrm{H}(\theta_p)\mathbf{R}_{(c)}^{-1}\mathbf{a}(\theta_p)=\frac{1}{\sigma^2_n}\Big(N-\frac{\arrowvert\mathbf{a}^H(\theta_p)\mathbf{a}_l\arrowvert^2}{\gamma+N} \Big).
\end{align}
Note that $ \|\mathbf{a}_l\|^2=\|\mathbf{a}(\theta_p)\|^2=N$. It should be mentioned that an analytical evaluation of the summation in (\ref{App Ripn}) may be difficult. A rough estimation is achieved when the inner product in (\ref{R4}) is approximated as
\begin{equation}
\arrowvert\mathbf{a}^H(\theta_p)\mathbf{a}_l\arrowvert^2\simeq
\begin{cases}
N^2, & \theta_p = \theta_l
\\
0, & \theta_p \neq \theta_l
\end{cases}
\end{equation}
It is assumed that the approximation is correct if, the angles $ \{ \theta_p\}_{p=1}^P $, are chosen in such a way that only one of them coincides with the interference direction, $ \theta_l $, (It means that the other angles, $\theta_p \neq \theta_l$ fall outside the main-beam of the function $ \arrowvert\mathbf{a}^\mathrm{H}(\theta_p)\mathbf{a}_l\arrowvert^2 $ ) and, the number of sensors, $ N $ is large enough. Hence, this approximation is utilized to compute the summation in \eqref{App Ripn} as follow \begin{align}
\mathbf{R}_{(c)\text{ipn}}\simeq&\frac{\sigma^2_n}{N}\sum_{\theta_p \neq \theta_l}\mathbf{a}(\theta_p)\mathbf{a}^H(\theta_p)\Delta\theta+\big(\frac{\sigma^2_n+N\sigma^2_l}{N} \big)\mathbf{a}_l\mathbf{a}_l^H\Delta\theta \nonumber\\ =&\frac{\sigma^2_n}{N}\sum_{\theta_p\in\Theta_{\text{ipn}}}\mathbf{a}(\theta_p)\mathbf{a}^H(\theta_p) \Delta \theta+\sigma^2_l\mathbf{a}_l\mathbf{a}_l^H\Delta\theta.
\end{align}
Since the size of the set $\Theta_s$ is much smaller than the size of $\Theta_{\text{ipn}}$ (measuring "size" in terms of sum of lengths of the intervals that compose the sets, i.e., the Borel measure), it can be shown that \cite{mohammadzadeh2019robust}
\begin{align}\label{App integral}
\int_{\Theta_{\text{ipn}}} \mathbf{a}(\theta_p)\mathbf{a}^H(\theta_p) d \theta_p \approxeq \int_{[-\pi,\pi]}\mathbf{a}(\theta_p)\mathbf{a}^H(\theta_p) d \theta_p=2\pi \mathbf{I}_N
\end{align}
so that the summation can also be approximated by (\ref{App integral}).  The same considerations about the size of $\Theta_s$ and $\Theta_{\text{ipn}}$ also allow us to approximate $\Delta\theta\approx \frac{2\pi}{N}$, resulting in
\begin{align}\label{R8}
\mathbf{R}_{(c)\text{ipn}}\simeq \frac{2 \pi}{N}\sigma^2_n\mathbf{I}_N+\frac{2\pi}{N}\sigma^2_l\mathbf{a}_l\mathbf{a}_l^\mathrm{H}=\frac{2\pi}{N}\mathbf{R}_{(c)}.
\end{align}
when the original NPICM in (\ref{App R}) is Compared with (\ref{R8}), it can be seen that the reconstruction matrix, $\mathbf{R}_{(c)\text{ipn}}$ only multiplies the true matrix, $\mathbf{R}_{(c)}$ by a factor $\dfrac{2\pi}{N}$.\\
The disadvantages of NPICM reconstruction based on Capon are due to the approximation of the integral \eqref{Ripn} of the rank one matrices $\mathbf{a}(\theta)\mathbf{a}^H(\theta)$ (weighted by the corresponding incident power from direction $\theta$) with a summation that requires a large number of computation to be able to synthesize powers from signals accurately \cite{mohammadzadeh2018modified, wang2016robust,somasundaram2011evaluation}. However, as we show in this paper, the proposed REC-DFT method estimates the NPICM without the need to estimate the interferences and the corresponding array SVs, resulting in an algorithm with an overall low complexity that is very competitive when compared with other methods in the literature.
\subsection{Proposed REC-DFT Approach}
\label{sec:2}

In order to achieve the optimal solution of the beamformer depicted in (\ref{optimal wegight vector}), we need to estimate the NPICM and the desired signal SV. In this section, we utilize a low-complexity spatial sampling process to reconstruct the robust adaptive REC-DFT beamformer with highly accurate SINR.
Let us consider the theoretical autocovariance sequence (ACS), \eqref{Theoretical R}, for single interference as
\begin{align} \label{R with single Interf.}
    \mathbf{R}= \sigma_s^2\mathbf{{a}}_s\mathbf{{a}}_s^H+\sigma_l^2\mathbf{{a}}_l\mathbf{{a}}_l^H+\sigma_n^2 \mathbf{I}=\mathbf{R}_s+\mathbf{R}_{\text{ipn}}
\end{align}
where $\mathbf{a}_s=[1, e^{-j\theta_s}, \cdots, e^{-j(N-1)\theta_s}]$ and $\mathbf{a}_l=[1, e^{-j\theta_l}, \cdots, e^{-j(N-1)\theta_l}]$. On the other hand, a covariance matrix is a Hermitian matrix with variances in the diagonal elements and covariances in the off-diagonal elements. If the signals $x_n(k)$ obtained at each antenna are gathered in the vector $\mathbf{x}(k)=[x_0(k) \ x_1(k) \cdots x_{N-1}(k)]^\mathrm{T}$ then the SCM  $\hat{\mathbf{R}}$, \eqref{SCM} can be written as follows
\begin{equation}
\hat{\mathbf{R}} =
\begin{bmatrix}
\hat{\mathbf{R}}(0,0) & \cdots & \hat{\mathbf{R}}(0,-(N-1)) \\
\hat{\mathbf{R}}(1,0) & \cdots & \hat{\mathbf{R}}(1,-(N-1)) \\
\vdots  & \ddots & \vdots  \\
\hat{\mathbf{R}}(N-1,0) & \cdots & \hat{\mathbf{R}}(N-1,-(N-1))
\end{bmatrix}
\end{equation}
This paper introduces a new approach for ULA (with identical spacing of sensors) in order to reconstruct the NPICM based on DFT coefficients of the auto correlation sequence of the measured data. Then, the power spectrum is estimated using the discrete Fourier transform. The whole spirit of the purpose is based on the ACS or the covariance function of the array observation vector. Thus, we need to obtain the correlation sequence of the received signal. This is achieved from the first column and first row of the SCM, where averaging the diagonals of this matrix improves the estimation of the correlation sequence as follows
\begin{equation} \label{correlation coefficients}
  \hat{r}(n) =\dfrac{1}{N-|n|}
    \begin{cases}
    \sum_{k=-n+1}^{N}\hat{\mathbf{{R}}}(k,n+k) & \text{$n<0$} \\ \\
      \sum_{k=1}^{N-n}\hat{\mathbf{{R}}}(k,n+k) & \text{$n \geqslant 0$}
    \end{cases}
\end{equation}
where the estimated correlation sequence of the  received signal is denoted as $ \hat{r}(n) $ for $ n=0,\pm1,...,\pm{(N-1)} $. Here, diagonal refers to not only the main diagonal, but all diagonals parallel to the main diagonal. The nth diagonal above the main diagonal may be expressed as  $[\hat{r}(k,k+n)]_{k=1, \cdots, N-n}, \text{for} \  n=1, \cdots, N-1$, and the nth diagonal below the main as $[\hat{r}(k,k-n)]_{k=n+1, \cdots, N}, \text{for} \  n=1, \cdots, N-1$.  It should be noted that the SCM is Hermitian. However, it is not Toeplitz in general, meaning that the averages of the diagonals below the main diagonal are not equal to the averages of the diagonals above it. Therefore, it is statistically more sound to use the first column as well as the first row.\\
By decomposing \eqref{correlation coefficients} as
\begin{align}
    [\hat{r}_s(n);& \  n=-(N-1), \cdots, (N-1)] \nonumber \\&\approx c \ \hat{\sigma}_s^2[e^{-j(N-1)\theta_s} \cdots, \ 1 \ , \cdots e^{j(N-1)\theta_s}] \nonumber \\&=c \ \hat{\sigma}^2_s [e^{j n \theta_s}; \ n=-(N-1), \cdots, (N-1)]
\end{align}
and
\begin{align}
    [\hat{r}_l(n);& \  n=-(N-1), \cdots, (N-1)] \nonumber \\&\approx c \ \hat{\sigma}_l^2[e^{-j(N-1)\theta_l} \cdots, \ 1 \ , \cdots e^{j(N-1)\theta_l}] \nonumber \\&=c \ \hat{\sigma}^2_l [e^{j n \theta_l}; \ n=-(N-1), \cdots, (N-1)]
\end{align}
then, we can write
\begin{align} \label{r hat}
    \hat{r}(n)\approx c \ (\hat{\sigma}^2_s e^{j n \theta_s}+ \hat{\sigma}^2_l e^{j n \theta_l}+\hat{\sigma}^2_n\delta_n)
\end{align}
where $c$ is the constant number which comes from taking average of every diagonal of SCM. Also, the spatial power spectrum of the received signal in terms of the autocorrelation coefficients and as a continuous function of the direction $\theta$ \cite{stoica2005spectral} is estimated as
\begin{align}\label{Phat}
\hat{P}(\theta)=\sum_{n=-(N-1)}^{N-1} \hat{r}(n)e^{-jn\theta},\hspace{.3cm}-\pi\le\theta\le\pi,
\end{align}
By replacing \eqref{r hat} into \eqref{Phat}, the spatial power spectrum can be derived as follows, \\
\begin{figure*}[h]
    \begin{align}\label{eq:power.spectrum}
    \hat{P}(\theta)&= \sigma^2_n+ \sum_{n=-(N-1)}^{N-1} \Big( \sigma^2_s e^{j n (\theta_s-\theta)}+ \sigma^2_l e^{j n (\theta_l-\theta)} \Big)  \nonumber \\ &= \sigma^2_n+ \sigma^2_s e^{-j(N-1)(\theta_s-\theta)} \sum_{n=0}^{2N-2} e^{j n (\theta_s-\theta)}+ \sigma^2_l e^{-j(N-1)(\theta_i-\theta)}  \sum_{n=0}^{2N-2} e^{j n (\theta_l-\theta)} \nonumber \\
    &=  \sigma^2_s \dfrac{\sin [(N-\frac{1}{2})(\theta_s-\theta)]}{\sin[ \frac{1}{2}(\theta_s-\theta)]}+ \sigma^2_l \dfrac{\sin [(N-\frac{1}{2})(\theta_l-\theta)]}{\sin[ \frac{1}{2}(\theta_l-\theta)]}+\sigma^2_n\nonumber \\ &=\hat{P}_s(\theta)+\hat{P}_{\text{ipn}}(\theta)
    \end{align}
    \noindent\rule{\textwidth}{.5pt}
\end{figure*}
In practice, calculating the spatial power spectrum, $ \hat{P}(\theta) $, over the entire range of directions $ (-\pi,\pi) $ is nearly impossible. Therefore, in order to compute $ \hat{P}(\theta) $, the direction variable must be sampled. If we take $ N_\text{DFT} $ samples in each period of $ \hat{P}(\theta) $, the spacing between angle points will be $ \Delta\theta=2\pi/N_\text{DFT} $. Therefore, the central angles of the set of bins that we obtain will be $ \theta_{i}=2\pi i/N_\text{DFT}$, where we can select $ i=(-N_\text{DFT}/2),...,(N_\text{DFT}/2)-1$.
$ N_\text{DFT} $ is selected in such a way that the array's angular resolution is larger than the angular span of a bin.\\
Moreover, it is known that, given the power spectrum, the autocorrelation sequence may be determined by taking the inverse discrete Fourier transform (IDFT) of $\hat{P}(\theta)$. However, from \eqref{eq:power.spectrum} it can be seen that the estimated power spectrum has two parts: the spatial power spectrum of the desired signal, $\hat{P}_s(\theta)$, and the noise-plus-interference section, $\hat{P}_{\text{ipn}}(\theta)$. Therefore, in order to compute the correlation sequence associated with the NPICM, it is needed to find the IDFT of the noise-plus-interference section while zeroing the power spectrum in the direction of the SOI as follows
\begin{align}\label{correlation sequence of IPNC}
\tilde{r}_{\text{ipn}}(n)=\dfrac{1}{N_\text{DFT}}\sum_{\theta_i \in\Theta_{\text{ipn}}}\hat{P}_{\text{ipn}}(\theta_{i})e^{jn\theta_{i}},
\end{align}
The angle bins, which are in the noise-plus-interference region, $ \Theta_{\text{ipn}} $, capture the dominant DFT coefficients. \\
For the sake of simplicity, we consider the sequence $\tilde{r}_{\text{ipn}}(n)=\tilde{r}_n$ and construct the corresponding $N \times N$ Toeplitz matrix $\tilde{\mathbf{R}}_\mathrm{ipn}=[\tilde{r}_{k-j}; k,j=0,1,\cdots, N-1]$ as
\begin{equation} \label{Reconstructed IPNC}
\tilde{\mathbf{R}}_{\text{ipn}} =
\begin{bmatrix}
\tilde{r}_0 & \tilde{r}_{-1} & \tilde{r}_{-2} &\cdots & \tilde{r}_{-(N-1)} \\
\tilde{r}_{1} &\tilde{r}_0 & \tilde{r}_{-1} & \cdots & \tilde{r}_{-(N-2)} \\
\tilde{r}_{2} &\tilde{r}_{-1} & \tilde{r}_{0} & \cdots & \tilde{r}_{-(N-3)}\\
\vdots  & \vdots & \vdots &\ddots & \vdots \\
\tilde{r}_{(N-1)} & \cdots & \cdots &\cdots & \tilde{r}_0
\end{bmatrix}
\end{equation}

Moreover, the most important issue with NPICM reconstruction is the accuracy of the power spectrum estimate. The Capon spectral estimator, employed by the NPICM re-construction based methods, is not very accurate due to the summation approximation of the integral. This approximation may not capture the spectrum depending on the choice of the angular grid (sampling angles). The DFT coefficients capture all the spectral components in the received signal (as far as the truncated ACS can reveal), whereas in the Capon estimate only the integral form can capture the spectrum. This leads to a more accurate representation of the power spectrum by the available autocorrelation matrix estimate (REC-DFT), and therefore a more accurate reconstruction of the NPICM.

\section{The Desired Signal SV Estimation}

In this section, we describe a simple method in which the actual SV is estimated based on the DSCM. To obtain a good estimate, we propose using a priori knowledge that the impinging angle of the desired signal is outside $\Theta_{\text{ipn}}$.

Using \eqref{Theoretical R}, the DSCM is estimated by subtracting the noise-plus-interference signal component from the SCM as follows:
\begin{align} \label{Rs}
 \tilde{\mathbf{{R}}}_s=  \hat{\mathbf{{R}}}-\tilde{\mathbf{{R}}}_{\text{ipn}}.
\end{align}
However, it is known that the estimated DSCM $ \tilde{\mathbf{{R}}}_{s} $, is contaminated by the white noise that are the residual components which can be expressed as
\begin{align}
    \tilde{\mathbf{{R}}}_{s}= \sigma^2_{n(\text{res})} \mathbf{{I}}+ \tilde{\sigma}^2_s \mathbf{{a}}_s \mathbf{{a}}_s^H,
\end{align}
where $ \sigma^2_{n(\text{res})} $ and $ \tilde{\sigma}^2_s $ are the residual noise and the desired signal power, respectively, while the SOI\textquoteright s SV is denoted by  $ \mathbf{{a}}_s $. The basic idea of the proposed desired signal SV estimation is based on  multiplication of the reconstructed DSCM and the assumed SV of the SOI, where it is estimated as follows
\begin{align}\label{atilda}
\tilde{\mathbf{a}}_s=\tilde{\mathbf{R}}_s \bar{\mathbf{a}}_s=&(\sigma^2_{n(\text{res})} \mathbf{{I}}+ \tilde{\sigma}^2_s \mathbf{{a}}_s \mathbf{{a}}_s^H)\bar{\mathbf{a}}_s \nonumber \\=&\sigma^2_{n(\text{res})}\bar{\mathbf{a}}_s+\tilde{\sigma}^{2}_s(\mathbf{{a}}^{H}_s \bar{\mathbf{a}}_s)\mathbf{a}_s,
\end{align}
where $\bar{\mathbf{a}}_s$ stands for the assumed SV of the desired signal. In \eqref{atilda} the residual term $\sigma^2_{n(\text{res})} \mathbf{I}$  represents the noise power that falls within the desired signal’s angular sector which has not been accounted for in $\tilde{\mathbf{R}}_{\text{ipn}}$, and therefore has not been subtracted from the total covariance matrix. Since the angular sector of the desired signal, $\Theta_s$, is much smaller than the whole DoA range of $2 \pi$, the noise power in this sector is much smaller than the total noise power. Hence, the norm of the residual term  $\sigma^2_{n(\text{res})} \| \bar{\mathbf{a}}_s \|^2$  can be expected to be much smaller than the power of the signal term $\tilde{\sigma}^2_s | \mathbf{a}_s^H \bar{\mathbf{a}}_s |$. This can be better understood by noting that $\| \bar{\mathbf{a}}_s \|^2= \| \mathbf{
a}_s \|^2= N$  for ideal form of the SV, and $|\mathbf{a}_s^H \bar{\mathbf{a}}_s | \approx N$  if $\bar{\mathbf{a}}_s$ is close to $\mathbf{a}_s$ . Then it is sufficient that $\sigma^2_{n(\text{res})} \ll N \tilde{\sigma}^2_s $, which can be satisfied even for low SNR values. The accuracy of the SV estimate \eqref{atilda} can be investigated by calculating the beamformer SINR using this estimate as follows.

In the derivation of the SINR for the beamformer based on the SV estimate in \eqref{atilda}, the assumption is made that the NPICM is exact (i.e. $\tilde{\mathbf{R}}_{\text{ipn}}=\mathbf{R}_{\text{ipn}}$ ). This assumption may be justified by noting that the exclusion of the desired signal angular sector in the reconstruction of $\mathbf{R}_{\text{ipn}}$ is negligible if this sector is much smaller than the total $2 \pi$ range. Then, the SINR becomes
\begin{align} \label{sinr as}
\mathrm{SINR}=\sigma^2_s \dfrac{| \tilde{\mathbf{a}}_s^H {\mathbf{{R}}}_{\text{ipn}}^{-1}{\mathbf{{a}}}_s |^2}{\tilde{\mathbf{{a}}}_s^H {\mathbf{{R}}}_{\text{ipn}}^{-1}\tilde{\mathbf{{a}}}_s}.
\end{align}
In the ideal case with $\tilde{\mathbf{a}}_s= \mathbf{a}_s$, the optimum SINR is given by
\begin{align}\label{eq:optimal.SINR}
\mathrm{SINR}_\text{opt}=\sigma^2_s | \mathbf{a}_s^H \mathbf{R}_{\text{ipn}}^{-1}{\mathbf{{a}}}_s |.
\end{align}
By direct substitution of \eqref{atilda} into \eqref{sinr as} and using the approximation $(1+x)^{-1} \approx 1-x$ for $|x| \ll 1$, we can write
\begin{align} \label{SINR as}
\mathrm{SINR}=&\sigma^2_s | \mathbf{a}_s^H \mathbf{R}_{\text{ipn}}^{-1}{\mathbf{{a}}}_s | \Big \lbrace 1- \epsilon^2 \nonumber \\ & \cdot \Big [ \dfrac{|\bar{\mathbf{a}}^H \mathbf{R}_{\text{ipn}}^{-1}\bar{\mathbf{a}}| | \mathbf{a}_s^H \mathbf{R}_{\text{ipn}}^{-1} \mathbf{a}_s|- | \bar{\mathbf{a}}^H \mathbf{R}_{\text{ipn}}^{-1} \mathbf{a}_s|^2}{|\bar{\mathbf{a}}^H \mathbf{a}_s|^2 | \mathbf{a}_s^H \mathbf{R}_{\text{ipn}}^{-1} \mathbf{a}_s|^2} \Big] \Big \rbrace,
\end{align}
where $\epsilon=\sigma^2_{n(\text{res})}/\tilde{\sigma}_s^2$ is assumed to be much less than one. Now, an insight into the dependence of the reduction in SINR on the error in the presumed SV $\bar{\mathbf{a}}_s$ can be obtained by considering the single interference case. In this scenario, we exploit the NPICM defined in \eqref{R with single Interf.} where $\mathbf{R}_{\text{ipn}}= \sigma_l^2\mathbf{{a}}_l\mathbf{{a}}_l^H+\sigma_n^2 \mathbf{I}$. The inverse of the NPICM can be obtained as \eqref{App R invers}. Therefore by utilizing this inversion, the numerator of the expression within the square brackets in \eqref{SINR as} becomes
\begin{align} \label{long formula}
    |\bar{\mathbf{a}}^H \mathbf{R}_{\text{ipn}}^{-1}\bar{\mathbf{a}}| | \mathbf{a}_s^H &\mathbf{R}_{\text{ipn}}^{-1} \mathbf{a}_s|- | \bar{\mathbf{a}}^H \mathbf{R}_{\text{ipn}}^{-1} \mathbf{a}_s|^2= \dfrac{1}{\sigma^4_n} \Big(N^2- |\bar{\mathbf{a}}^H \mathbf{a}_s|\nonumber \\&+\dfrac{2}{N+\gamma} \Re \Big \lbrace(\bar{\mathbf{a}}^H\mathbf{a}_s) (\mathbf{a}_s^H \mathbf{a}_l) (\mathbf{a}_l^H \bar{\mathbf{a}}) \Big \rbrace \nonumber \\ & - \dfrac{N}{N+\gamma}(|\bar{\mathbf{a}}^H \mathbf{a}_l |^2+ |\mathbf{a}^H_s \mathbf{a}_l|^2) \Big ),
\end{align}
where $\Re \lbrace \cdot \rbrace$ indicates the real part of a complex number. If the SVs have the ideal form, it can be easily shown that
\begin{align}
    \bar{\mathbf{a}}_s^H \mathbf{a}_s=&\sum_{q=0}^{N-1} e^{-j q (\theta_s-\bar{\theta})}=\dfrac{1-e^{jN(\bar{\theta}-\theta_s)}}{1-e^{j(\bar{\theta}-\theta_s)}} \nonumber \\ =&\dfrac{e^{j\frac{N}{2}(\bar{\theta}-\theta_s)}\Big(2j \sin (\frac{N}{2}(\bar{\theta}-\theta_s)) \Big ) }{e^{j\frac{1}{2}(\bar{\theta}-\theta_s)}\Big(2j \sin (\frac{1}{2}(\bar{\theta}-\theta_s)) \Big )},
\end{align}
and then we can write
\begin{align} \label{sinc}
  |\bar{\mathbf{a}}_s^H \mathbf{a}_s|^2= \dfrac{\sin^2 \big(\frac{N}{2}(\bar{\theta}-\theta_s)\big )}{\sin^2 \big(\frac{1}{2}(\bar{\theta}-\theta_s)\big)}.
\end{align}
where $\bar{\theta}$ is the DoA corresponding to the presumed SV, $\bar{\mathbf{a}}_s$.
Using the following Taylor series expansion of the rhs of \eqref{sinc} with $\Phi=\bar{\theta}-\theta_s$ , assumed to be much smaller than one, we can write that
\begin{align} \label{norm of as and as bar}
   |\bar{\mathbf{a}}_s^H \mathbf{a}_s|^2=\dfrac{\sin^2(N \Phi/2)}{\sin^2(\Phi/2)} \approx N^2- \dfrac{1}{24}N^2(N^2-1)\Phi^2.
\end{align}
Now, if the interference DoA is sufficiently separated from the DoAs of $\bar{\mathbf{a}}_s$ and $\mathbf{a}_s$, the contributions of the terms in \eqref{long formula} involving the interference SV $\mathbf{a}_l$ become negligible compared to the first two terms. Thus, \eqref{long formula} can be re-written as
\begin{align} \label{summarized long formula}
    |\bar{\mathbf{a}}_s^H \mathbf{R}_{\text{ipn}}^{-1}\bar{\mathbf{a}}_s| | \mathbf{a}_s^H \mathbf{R}_{\text{ipn}}^{-1} \mathbf{a}_s|&- | \bar{\mathbf{a}}_s^H \mathbf{R}_{\text{ipn}}^{-1} \mathbf{a}_s|^2 \nonumber \\ &\approx \dfrac{1}{\sigma^4_n} \Big(N^2- |\bar{\mathbf{a}}_s^H \mathbf{a}_s|^2 \Big ).
\end{align}
By exploiting \eqref{norm of as and as bar} and replacing \eqref{summarized long formula} in \eqref{SINR as}. we have
\begin{align}\label{eq:approx.SINR}
    \mathrm{SINR} \approx \sigma^2_s | \mathbf{a}_s^H \mathbf{R}_{\text{ipn}}^{-1}{\mathbf{{a}}}_s | \big( 1- \dfrac{1}{12} \epsilon^2 \Phi^2 \big ).
\end{align}
The residual noise power in the desired signal angular sector may be taken to be proportional to the width of this sector, assuming that the noise is spatially white. Then, the residual noise power can be expressed as
\begin{align}
    \sigma^2_{n(\text{res})}= \dfrac{N_s}{N_\text{DFT}} \sigma^2_n
\end{align}
where $N_s$ is the number of frequency bins in the desired signal angular sector. Hence, we have
\begin{align}
    \epsilon=\dfrac{\sigma^2_{n(\text{res})}}{\tilde{\sigma}^2_s} \approx \dfrac{N_s}{N_\text{DFT} \times \text{SNR}}
\end{align}
Since, $N_s \ll N_\text{DFT}$, $\epsilon$ may be expected to be much smaller than one even for low SNR values. \\
\indent Comparing \eqref{eq:optimal.SINR} with \eqref{eq:approx.SINR}, it is observed that the SV $ \tilde{\mathbf{a}}_s $ is a good estimate for the desired signal SV.\\
Using the corrected SV of SOI (\ref{atilda}), $ \tilde{\mathbf{a}}_s $ and NPICM \eqref{Reconstructed IPNC} into (\ref{optimal wegight vector}), the weight vector of the proposed beamformer is given as
\begin{align}\label{weight}
\mathbf{{w}}_{\text{prop}}=\dfrac{\tilde{\mathbf{{R}}}_{\text{ipn}}^{-1}\tilde{\mathbf{{a}}}_s}{\tilde{\mathbf{{a}}}_s^H\tilde{\mathbf{{R}}}_{\text{ipn}}^{-1}\tilde{\mathbf{{a}}}_s}
\end{align}

Algorithm 1 summarizes the steps to obtain the proposed adaptive beamforming weights.

\begin{algorithm}
    \caption{Proposed REC-DFT Adaptive Beamforming }\label{sobelcode}
    1: \textbf{Input:}\:Array received data vector $\lbrace \mathbf{{x}}(k) \rbrace_{k=1}^K $,\\
    2: \text{Initialize:}\: Compute the SCM\\  ${\hspace{1em}}\hat{\mathbf{{R}}}=(1/K)\sum_{k=1}^{K} \mathbf{{x}}(k)\mathbf{{x}}^H(k)$; \ $N_\text{DFT}=38$; \ $\Delta \theta= \frac{2\pi}{N_\text{DFT}}$; \\
    3: \text{For} $ n=-(N-1):(N-1)$ \\
    4:   \qquad    \text{For} $ k=1:N-n$\\
    5:  \qquad  $\hat{r}(n)=\dfrac{1}{N-|n|}\sum_{k=1}^{N-|n|}\hat{\mathbf{{R}}}(k,n+k)$\\
    6: \qquad \text{End For}\\
    7: \text{End For} \\
    8: \text{For} $i=1:\Delta \theta:N_\text{DFT}$\\
    9: \quad $\theta(i)=(-\pi/2)+(i-1)(\pi/N_\text{DFT})$\\
    10: \quad \text{If} \ $ \theta(i)\in \Theta_{\text{ipn}}  $ \quad \text{then}\\
    11: \qquad \quad  \text{For} $n=1: \mathrm{length}(n) $\\
    12:  \qquad \qquad   $\hat{P}(\theta(i))=\sum_{n=-(N-1)}^{N-1} \hat{r}(n)e^{-jn\theta(i)}$\\
    13: \qquad \quad \text{End For}\\
    14: \quad \text{End If}\\
    15: \text{End For}\\
    16: Compute correlation sequence, $\tilde{r}_{\text{ipn}}(n)=\tilde{r}_n$, using \eqref{correlation sequence of IPNC}\\
    17: Construct the NPICM based on $\tilde{\mathbf{R}}_\mathrm{ipn}=[\tilde{r}_{k-j}; k,j=0,1,\cdots, N-1]$ \\
    18: Estimate the DSCM as  $\tilde{\mathbf{{R}}}_s= \hat{\mathbf{{R}}}-\tilde{\mathbf{{R}}}_{\text{ipn}}$\\
    19: Calculate estimation of the desired signal SV $\tilde{\mathbf{{a}}}_s=\tilde{\mathbf{{R}}}_s \bar{\mathbf{a}}_s$\\
    20: Design proposed beamformer using \eqref{weight}\\
    21: \textbf{Output:}\: Proposed beamforming weight vector $ \mathbf{w}_{\text{prop}} $\\
\end{algorithm}

\section{Computational complexity}

We compare the computational complexities of the different methods, as summarized in Table~\ref{tab:expcond}. Our main contribution consists in developing a fast and numerically stable technique for NPICM reconstruction and the SOI steering vector estimation. Given an array of $N$ elements, $K$ (number of snapshots) and $Q$ (the number of uniform samples in the noise-plus-interference angular sector), the computational complexity for computing the NPICM \eqref{Reconstructed IPNC} is $ \mathcal{O}(N_\text{DFT}N^2)$ where $N_\text{DFT}=Q=38$ and the SOI steering vector estimation needs $ \mathcal{O}(KN^2)$ with $K \leq N \le Q$, the overall complexity of REC-DFT is $ \mathcal{O}(QN^2)$.\\
In \cite{huang2012modified}, the shrinkage method is used to compute the covariance matrix, which has a complexity of $\mathcal{O}(KN)$, whereas the SV estimation has a complexity of $\mathcal{O}(N^3)$. As a result, the total complexity is $ \mathcal{O}(N^3)$. The beamforming method in \cite{zhang2016interference} has a complexity of $ \mathcal{O}(QN^3)$. The beamformers in \cite{zheng2018covariance,chen2015robust} have a complexity of $ \mathcal{O}(QN^2)$ to reconstruct the NPICM, while the complexity of the beamformer in \cite{zheng2018covariance} to estimate the desired signal SV is dominated based on solving a quadratically constrained quadratic programme (QCQP) which is $\mathcal{O}(N^{3.5})$. The beamformer in \cite{mohammadzadeh2020maximum}, needs $ \mathcal{O}(QN^2)$  and $ \mathcal{O}(SN^2)$ complexity for NPICM reconstruction and the SV estimation, respectively, where $S$ is the number of sampling points of the desired signal angular sector. The computational complexity for the reconstruction of the signal covariance matrix for the beamformer in \cite{gu2019adaptive} is higher, due to the matrix inversion and eigen-decomposition $ \mathcal{O}(SN^3)$. The only algorithm with comparable complexity $ \mathcal{O}(N^3)$ is \cite{huang2012modified}, but the performance of the proposed algorithm is considerably superior, as shown in the next section.

The conclusion, when comparing the methods based on the computational complexity in Table~\ref{tab:expcond}, is that the proposed REC-DFT method has significantly lower complexity as compared to the other beamformers (while the performance is similar or better, as shown in the next section).\\
It is worth noting that, since some references did not provide values for the parameter of $Q$ and $S$, the values used for the tested beamformers and the proposed REC-DFT method given in Table~\ref{tab:expcond} are the lowest quantities that result in the best performance for each algorithm in the simulations.
\begin{table}[h] \label{com}
  \centering
  \caption{Computational Complexity}
  \label{tab:expcond}
  \begin{tabular}{c c c c c}
    \hline
    Beamformers  &Complexity &$N$  & $Q$  & $S$   \\
    \hline
REC-DFT (Proposed)  &$ \mathcal{O}(QN^2)$ &30 &38  &-  \\
REC-ES \cite{huang2012modified} & $ \mathcal{O}(N^3)$   &30 &-  &-  \\
REC-SPSS \cite{zhang2016interference} &$ \mathcal{O}(QN^3)$  &30 &30  &-  \\
REC-Re \cite{zheng2018covariance}  &$ \mathcal{O}(\max(QN^2,N^{3.5}))$  &30  & 300  &-  \\
REC-MEPS \cite{mohammadzadeh2020maximum}  & $\mathcal{O}(\max(SN^2,QN^2))$   &30 & - &20  \\
REC-CC \cite{chen2015robust}  &$ \mathcal{O}(QN^2)$  &30  &300  &-  \\
REC-OS \cite{gu2019adaptive} &$ \mathcal{O}(SN^3)$  &30  &-  &20  \\
    \hline
\end{tabular}
\end{table}
\section{Simulations}

In this section, the performance of the proposed REC-DFT algorithm is validated by computer simulations. We consider, a ULA with $N=30$ sensors with half-wavelength spaced. A complex  additive white Gaussian noise in each sensor, is modeled as a zero mean and unity variance. The signal sources consist of one SOI and two interferers where the nominal SOI direction is kept at $\bar{\theta}_s=0^o$, and the nominal interference DoAs are fixed at $ \lbrace \bar{\theta}_1, \bar{\theta}_2 \rbrace= \lbrace40^o, -50^o\rbrace$. For each interfering signal, the interference-to-noise ratio (INR) in a single sensor is equal to 30dB. In each scenario, 100 independent simulation runs have been performed to show the results. The SNR is fixed at 10dB when comparing the performance of adaptive beamforming algorithms in terms of the number of snapshots. The number of snapshots is fixed to be $K = 30$ when comparing the performance versus SNR. \\
The proposed REC-DFT beamformer was compared with the covariance matrix reconstruction based on the Capon spectrum and the SV estimation presented in \cite{chen2015robust} (REC-CC), the covariance matrix reconstruction based on maximum Entropy presented in \cite{mohammadzadeh2020maximum} (REC-MEPS), the covariance matrix reconstruction and the SV estimation presented in \cite{huang2012modified} (REC-ES), the covariance matrix reconstruction based on interference SV estimation presented in \cite{zheng2018covariance} (REC-Re), the covariance matrix reconstruction based on spatial power spectrum sampling presented in \cite{zhang2016interference} (REC-SPSS) and the covariance matrix reconstruction based on orthogonal subspace in \cite{gu2019adaptive} (REC-OS).\\
The SOI's angular sector and the interference are set as $ {\Theta}_s=[{\bar{\theta}_s}-4^\circ,{\bar{\theta}_s}+4^\circ] $  $ \Theta_{\text{ipn}}=[-90^\circ,{\bar{\theta}_s}-4^\circ)\cup({\bar{\theta}_s}+4^\circ,90^\circ] $, respectively. In the beamformer REC-Re, the upper bound of the norm of the SV mismatch is set to $ \epsilon=\sqrt{0.1} $. For the eigenspace-based beamformer \cite{huang2012modified} (REC-ES), the energy percentage is set as $\rho=0.9$. In the REC-SPSS beamformer, $\Delta=\sin^{-1}(2/N)$ is used to find the width of the dithering area. To solve all convex optimization problems the Matlab CVX toolbox \cite{grant2008cvx} is used.

\subsection{Mismatch Due to Random signal look direction}

\begin{figure}[!]
    \centering
    \includegraphics[width=3.5in]{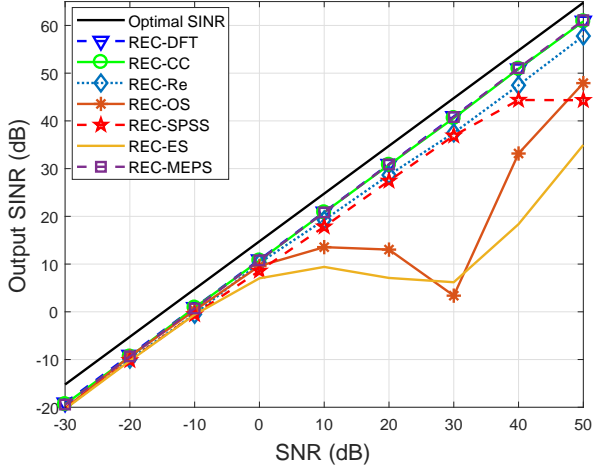}
    \caption{Output SINR versus SNR in case of random look direction error}
    \label{SNR_Look}
\end{figure}

\begin{figure}[!]
    \centering
    \includegraphics[width=3.5in]{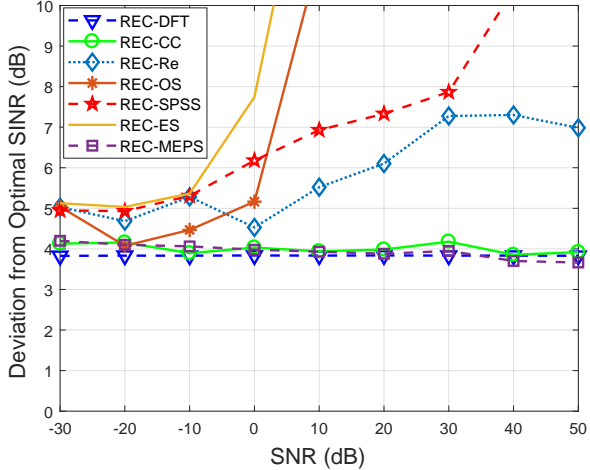}
    \caption{Deviation from optimal SINR versus SNR in case of random look direction error}
    \label{DEV_Look}
\end{figure}

\begin{figure}[!]
    \centering
    \includegraphics[width=3.5in]{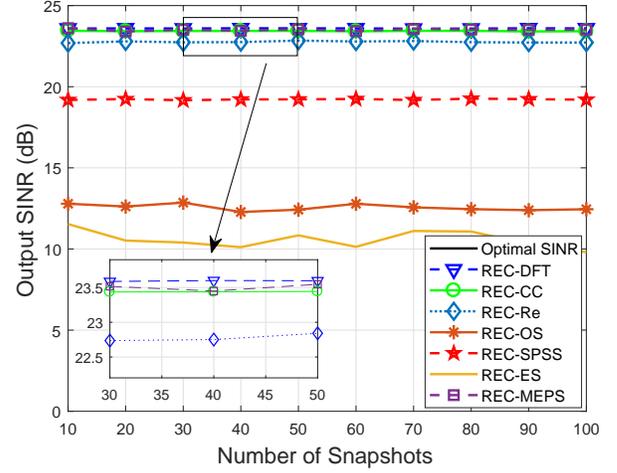}
    \caption{Output SINR versus number of snapshots in case of random look direction error}
    \label{Snap_Look}
\end{figure}

We consider a scenario with random look direction error in the first example. For each simulation run, it is assumed that DoA of the interferences and the desired signal follow a uniform distribution in $ [-4^o,4^o] $. It means that the direction remains fixed from snapshot to snapshot while varies from run to run. That is, the DoAs of two interferences are varied in $[36^o,44^o]$ and $[-54^o,-46^o]$ and  the actual DoA of the SOI is uniformly distributed in $ [-4^o,4^o] $. \\
The output SINR of the tested methods versus the input SNR is shown in Fig.~\ref{SNR_Look}. Since the proposed beamformer has essentially the same performance as \cite{zheng2018covariance}, \cite{chen2015robust} and \cite{mohammadzadeh2020maximum}, their deviations from the optimal SINR are shown in Fig.~\ref{DEV_Look}. It shows that the proposed REC-DFT beamformer has almost the same performance as the REC-MEPS and REC-CC beamformers. Since the proposed method avoids reconstructing the NPICM using the estimated SVs and corresponding powers, then it achieves almost near-optimal SINR for a large range of SNRs. Fig.~\ref{Snap_Look} depicts the output SINR of the tested beamformers versus the snapshots.
Comparing the output SINRs between the proposed method REC-DFT and the other methods, it can be seen that the proposed method achieves similar robustness against signal direction error, but with a substantially lower complexity for the computation of the NPICM.

\subsection{Mismatch Due to Incoherent Local Scattering} \label{Incoherent}
\begin{figure}[!]
    \centering
    \includegraphics[width=3.5in]{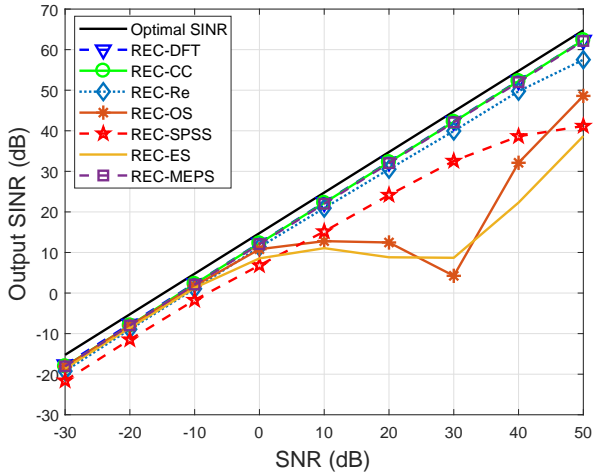}
    \caption{Output SINR versus SNR in case of Incoherent scattering}
    \label{SNR_Inco}
\end{figure}

\begin{figure}[!]
    \centering
    \includegraphics[width=3.5in]{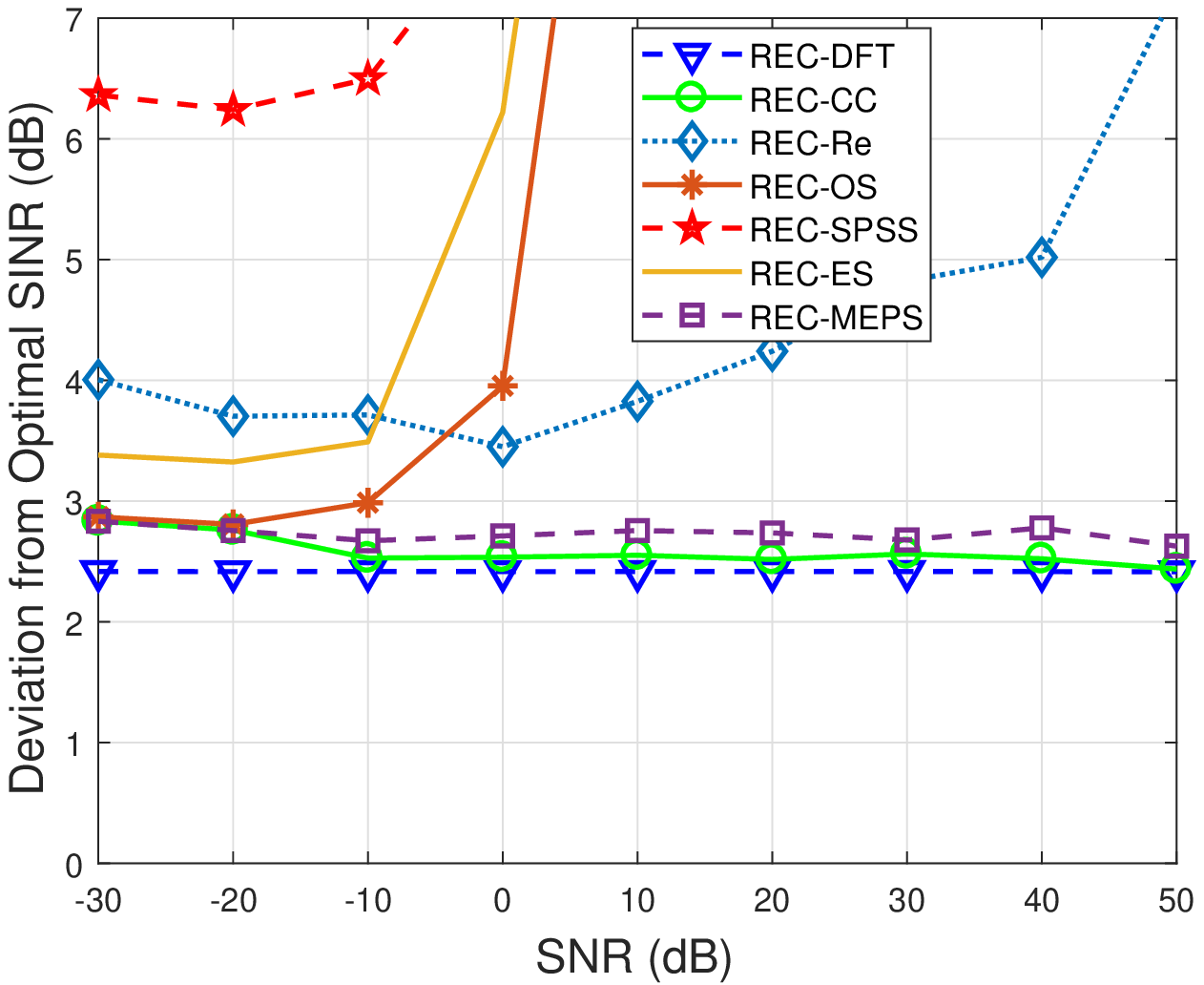}
    \caption{Deviation from optimal SINR versus SNR in case of Incoherent scattering}
    \label{DEV_Inco}
\end{figure}

\begin{figure}[!]
    \centering
    \includegraphics[width=3.5in]{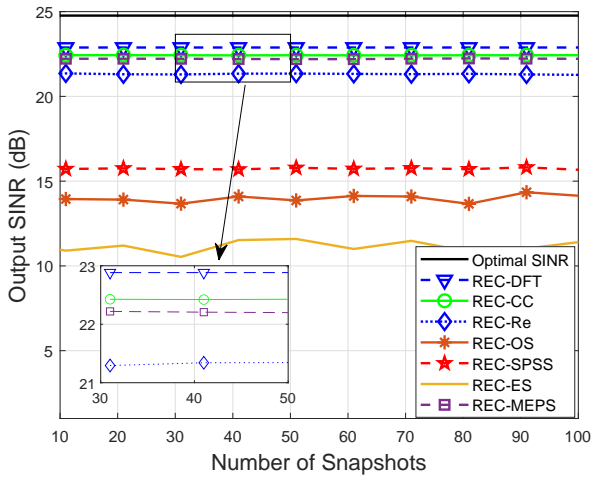}
    \caption{Output SINR versus number of snapshots in case of Incoherent scattering}
    \label{Snap_Inco}
\end{figure}

In the second example, a scenario with an incoherent local scattering of the desired signal is considered. The signal is assumed to have a time-varying spatial signature, and the SV of the desired signal is modeled as
\begin{align}
    \mathbf{{a}}(k)=s_0(k)\mathbf{{a}}(\bar{\theta}_s)+\sum_{r=1}^4 s_r(k)\mathbf{a}(\theta_r),
\end{align}
where $s_0(k)$ and $s_r(k)$ $(r=1,2,3,4)$ are independent and identically subject to the complex Gaussian distribution with zero mean. The DoAs $\{\theta_r\}$ are independently drawn in each simulation run from Gaussian distribution with $\bar{\theta}_s$ mean and standard deviation $ 2^o$. Note that $\theta_r$ and $\bar{\theta}_s$ vary from run to run while keeping unchanged from snapshot to snapshot. Simultaneously, the random variables  $s_0(k)$ and $s_r(k)$ change both from run to run and snapshot to snapshot. This is the scenario of incoherent local scattering \cite{besson2000decoupled}, where the DSCM $\mathbf{R}_s$ is not a rank-one matrix anymore and the output SINR should be rewritten in a general form
\begin{align}
\mathrm{SINR}_\mathrm{opt}=\dfrac{\mathbf{{w}}^H \mathbf{{R}}_{s}\mathbf{{w}} }{\mathbf{{w}}^H \mathbf{{R}}_{i+n}\mathbf{{w}}},
\end{align}
which is maximized by the weighted vector \cite{gershman1999robust}:
\begin{align}
    \mathbf{{w}}_\mathrm{opt}= \mathcal{P} \lbrace \mathbf{{R}}_{i+n}^{-1} \mathbf{{R}}_s \rbrace
\end{align}
where $\mathcal{P} \lbrace \cdot \rbrace$ represents the principal eigenvector of a matrix. Fig.~\ref{SNR_Inco} and Fig.~\ref{DEV_Inco} depict the output SINR of the tested beamformers versus the SNR and the deviation of the tested method from the optimal SINR respectively. The superior performance is due to the accurate estimation of the desired signal SV and NPICM. In order to reconstruct the DSCM, we use a low-complexity algorithm to remove the interference and noise components from the sample covariance matrix. The eigenvector corresponding to the largest eigenvalue in DSCM contains the most information which is achieved by multiplication of the assumed SV. The obtained SV is referred to as the desired signal SV.
The output SINR versus the number of snapshots is plotted in Fig.~\ref{Snap_Inco}. It is seen that the REC-DFT has higher accuracy SINR for SNR less that -10 dB compared to the REC-CC and REC-MEPS beamformers. Also, it is seen that the performance of the proposed method (REC-DFT) is almost the same as that of REC-CC for SNRs larger than 0 dB. {However the proposed REC-DFT method requires lower computational cost.}


\subsection{Array SV Error Due to wavefront distortion}

\begin{figure}[!]
    \centering
    \includegraphics[width=3.5in]{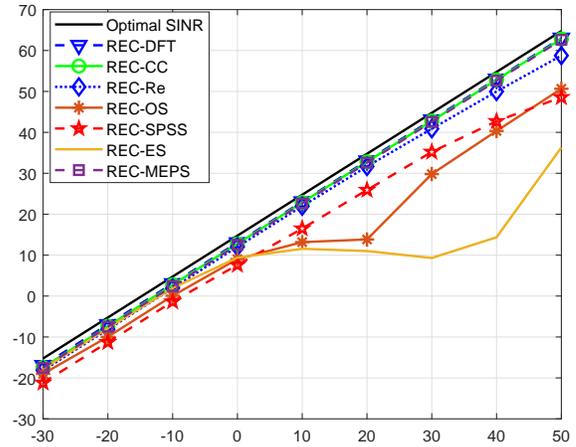}
    \caption{Output SINR versus SNR in case of wavefront distortion}
    \label{SNR_Wave}
\end{figure}

\begin{figure}[!]
    \centering
    \includegraphics[width=3.5in]{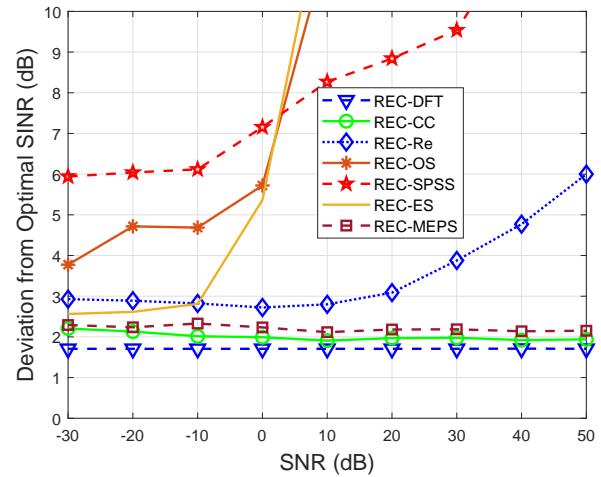}
    \caption{Deviation from optimal SINR versus SNR in case of wavefront distortion}
    \label{DEV_Wave}
\end{figure}

\begin{figure}[!]
    \centering
    \includegraphics[width=3.5in]{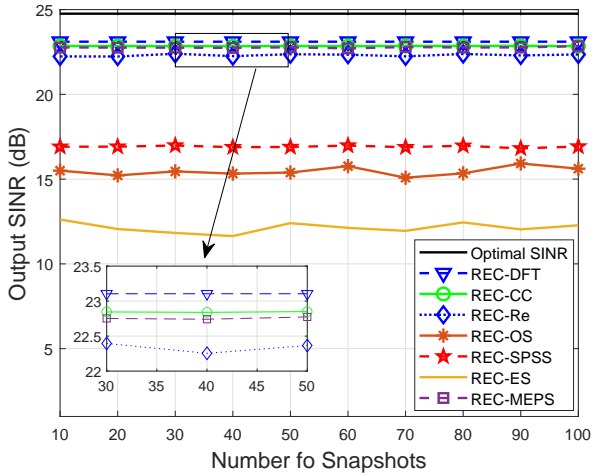}
    \caption{Output SINR versus number of snapshots in case of wavefront distortion}
    \label{Snap_Wave}
\end{figure}

In this example, a scenario is considered that the desired signal SV is distorted by wave propagation in which the medium is not uniform in character or content (inhomogeneous). The mismatch specifically states that the components of the presumed SV stack up the incremented phase of distortion independently. Assuming that in each simulation run, the phase increments are fixed and are independently chosen from a Gaussian random generator with zero mean and standard deviation of 0.04.\\
Fig.~\ref{SNR_Wave} shows the output SINR of the beamformers versus the input SNR. Fig.~\ref{DEV_Wave} illustrates the deviation from optimal output SINR versus the input SNR. It is distinct from the figures that the proposed beamformer achieves the better performance while the REC-CC and REC-MEPS demonstrate the acceptable performance against the wavefront distortion. The performance of the output SINR of all tested methods versus the number of snapshots is given in Fig.~\ref{Snap_Wave}. Similar to the previous scenarios, the proposed beamformer keeps its performance against mismatch which demonstrates the higher accuracy of the desired signal SV estimation and the NPICM reconstruction.

\subsection{Coherent Local Scattering Error}
\begin{figure}[!]
    \centering
    \includegraphics[width=3.5in]{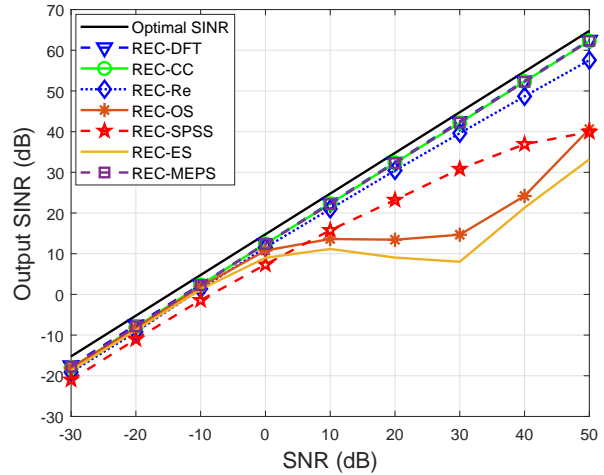}
    \caption{Output SINR versus SNR in case of coherent scattering}
    \label{SNR_Loc}
\end{figure}

\begin{figure}[!]
    \centering
    \includegraphics[width=3.5in]{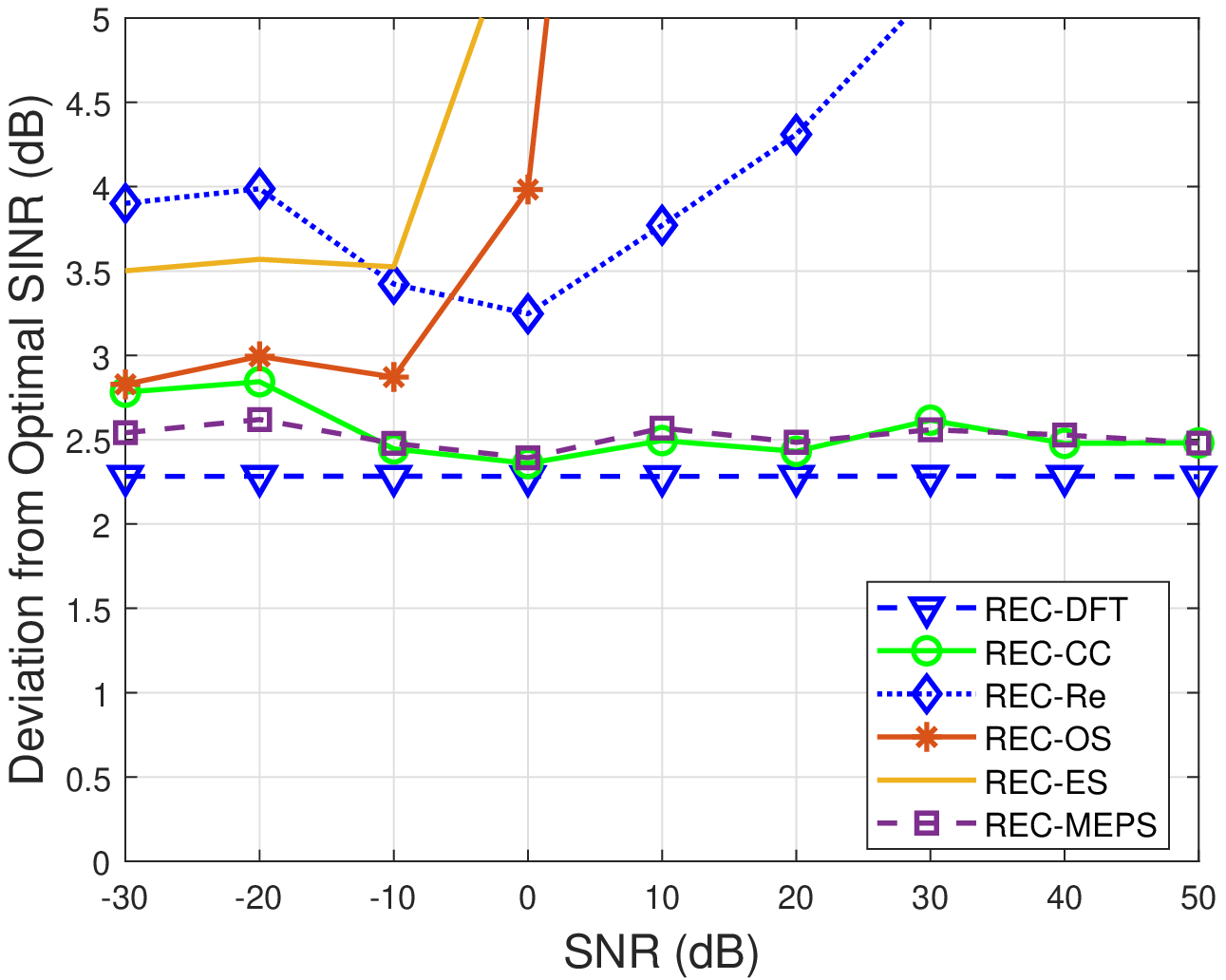}
    \caption{Deviation from optimal SINR versus SNR in case of coherent scattering}
    \label{DEV_Loc}
\end{figure}

\begin{figure}[!]
    \centering
    \includegraphics[width=3.5in]{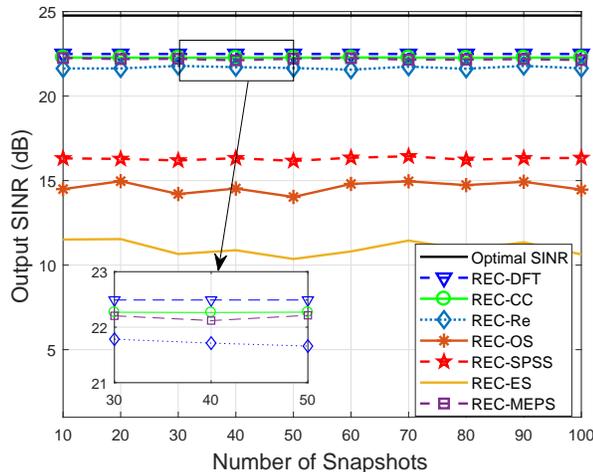}
    \caption{Output SINR versus number of snapshots in case of coherent scattering}
    \label{Snap_Loc}
\end{figure}

In the fourth example, we investigate a scenario in which the assumed signal array is a plane wave impinging from $ \bar{\theta}_s=0^\circ$, whereas the actual SV of SOI is composed of five signal paths as
\begin{align}
\hat{\mathbf{a}}_s=\bar{\mathbf{a}}_s+\sum_{i=1}^{4}e^{j\varphi_{i}} \mathbf{{b}}(\theta_{i}).
\end{align}
The SV corresponding to the direct path is denoted as $\bar{\mathbf{a}}_s$ and $ \{\theta_i \} $, represents the $i^{th}$ coherently scattered path which are generated in the same manner as the scenario in section \eqref{Incoherent}. In each simulation run, the parameter $ \{\varphi_{i}\} $ depicts the path phases that are drawn uniformly from the interval $ [0,2\pi]$. Also, $  \{\theta_i \} $ and $ \{\varphi_{i}\} $  vary from run to run while keeping unchanged from snapshot to snapshot. Note that the SNR in this example is defined by taking into account all signal paths.\\
In Fig.~\ref{SNR_Loc}, the performance versus SNR with a fixed number of snapshots is shown. Furthermore, the performance difference of the proposed beamformer and REC-MEPS and REC-CC is compared in Fig.~\ref{DEV_Loc}. It is shown that the proposed method outperforms other tested beamformers in terms of robustness against local scattering mismatch, which is primarily due to a more accurate estimation of the NPICM and SV of the SOI. We can see that the REC-SPSS beamformer has nearly the same performance as the other scenarios because this method only integrates the interference region rather than the signal region. The performance of the algorithm will remain unchanged as long as the interference region does not contain the SOI. {In the REC-CC beamformer the performance is degraded since it can not eliminate the subspace swap error in the case of low SNRs.}  Other algorithms, on the other hand, clearly suffer significant performance losses because the SV of the desired signal is influenced by the local scattering error.\\
Also, Fig.~\ref{Snap_Loc} shows the performance of the proposed REC-DFT method with fixed SNR while the number of snapshots $ K $ is changed. We notice that the proposed REC-DFT beamformer is also robust against this kind of uncertainty for whole range of the snapshots.

\section{Conclusion}
In this paper, we have introduced REC-DFT to reconstruct the IPNC matrix based on the idea of reconstructing the autocorrelation sequence of a random process from a set of measured data. The DFT of the correlation sequence is employed to estimate the power spectrum of the signals. A significant advantage of the proposed robust REC-DFT adaptive beamforming is that only little prior information is required. An imprecise knowledge of the angular sectors in which the interferences are located is sufficient for the proposed REC-DFT algorithm. Simulation results demonstrate that the proposed REC-DFT method has excellent performance, in many situations superior to that of other methods, while requiring lower computational complexity.

\section*{Acknowledgment}

This work was supported in part by the São Paulo Research
Foundation (FAPESP) through the ELIOT project under Grant
2018/12579-7 and Grant 2019/19387-9.

\bibliographystyle{IEEEtran}
\bibliography{mybibfile}

\begin{thebibliography}{10}
\providecommand{\url}[1]{#1}
\csname url@samestyle\endcsname
\providecommand{\newblock}{\relax}
\providecommand{\bibinfo}[2]{#2}
\providecommand{\BIBentrySTDinterwordspacing}{\spaceskip=0pt\relax}
\providecommand{\BIBentryALTinterwordstretchfactor}{4}
\providecommand{\BIBentryALTinterwordspacing}{\spaceskip=\fontdimen2\font plus
\BIBentryALTinterwordstretchfactor\fontdimen3\font minus
  \fontdimen4\font\relax}
\providecommand{\BIBforeignlanguage}[2]{{%
\expandafter\ifx\csname l@#1\endcsname\relax
\typeout{** WARNING: IEEEtran.bst: No hyphenation pattern has been}%
\typeout{** loaded for the language `#1'. Using the pattern for}%
\typeout{** the default language instead.}%
\else
\language=\csname l@#1\endcsname
\fi
#2}}
\providecommand{\BIBdecl}{\relax}
\BIBdecl

\bibitem{van2004detection}
L.~Harry and V.~Trees, \emph{Optimum Array Processing: part IV of Detection,
  Estimation, and Modulation Theory}.\hskip 1em plus 0.5em minus 0.4em\relax
  John Wiley and Sons, New York, 2002.

\bibitem{rahmani2014robust}
M.~Rahmani and M.~H. Bastani, ``Robust and rapid converging adaptive
  beamforming via a subspace method for the signal-plus-interferences
  covariance matrix estimation,'' \emph{IET Signal Processing}, vol.~8, no.~5,
  pp. 507--520, 2014.

\bibitem{liu2018coprime}
K.~Liu and Y.~D. Zhang, ``Coprime array-based robust beamforming using
  covariance matrix reconstruction technique,'' \emph{IET Communications},
  vol.~12, no.~17, pp. 2206--2212, 2018.

\bibitem{vorobyov2003robust}
S.~A. Vorobyov, A.~B. Gershman, and Z.~Q. Luo, ``Robust adaptive beamforming
  using worst-case performance optimization: A solution to the signal mismatch
  problem,'' \emph{IEEE Transactions on Signal Processing}, vol.~51, no.~2, pp.
  313--324, 2003.

\bibitem{rdrcb}
S.~D. Somasundaram, N.~H. Parsons, P.~Li, and R.~C. de~Lamare,
  ``Reduced-dimension robust capon beamforming using krylov-subspace
  techniques,'' \emph{IEEE Transactions on Aerospace and Electronic Systems},
  vol.~51, no.~1, pp. 270--289, 2015.

\bibitem{jidf}
R.~C. de~Lamare and R.~Sampaio-Neto, ``Adaptive reduced-rank processing based
  on joint and iterative interpolation, decimation, and filtering,'' \emph{IEEE
  Transactions on Signal Processing}, vol.~57, no.~7, pp. 2503--2514, 2009.

\bibitem{jio}
------, ``Reduced-rank adaptive filtering based on joint iterative optimization
  of adaptive filters,'' \emph{IEEE Signal Processing Letters}, vol.~14,
  no.~12, pp. 980--983, 2007.

\bibitem{jiols}
------, ``Reduced-rank space–time adaptive interference suppression with
  joint iterative least squares algorithms for spread-spectrum systems,''
  \emph{IEEE Transactions on Vehicular Technology}, vol.~59, no.~3, pp.
  1217--1228, 2010.

\bibitem{sjidf}
R.~Fa, R.~C. de~Lamare, and L.~Wang, ``Reduced-rank stap schemes for airborne
  radar based on switched joint interpolation, decimation and filtering
  algorithm,'' \emph{IEEE Transactions on Signal Processing}, vol.~58, no.~8,
  pp. 4182--4194, 2010.

\bibitem{l1stap}
Z.~Yang, R.~C. de~Lamare, and X.~Li, ``$l_1$ -regularized stap algorithms with
  a generalized sidelobe canceler architecture for airborne radar,'' \emph{IEEE
  Transactions on Signal Processing}, vol.~60, no.~2, pp. 674--686, 2012.

\bibitem{rrdstap}
X.~Wang, Z.~Yang, J.~Huang, and R.~C. de~Lamare, ``Robust two-stage
  reduced-dimension sparsity-aware stap for airborne radar with coprime
  arrays,'' \emph{IEEE Transactions on Signal Processing}, vol.~68, pp. 81--96,
  2020.

\bibitem{jiolcmv}
\BIBentryALTinterwordspacing
R.~{de Lamare}, L.~Wang, and R.~Fa, ``Adaptive reduced-rank lcmv beamforming
  algorithms based on joint iterative optimization of filters: Design and
  analysis,'' \emph{Signal Processing}, vol.~90, no.~2, pp. 640--652, 2010.
  [Online]. Available:
  \url{https://www.sciencedirect.com/science/article/pii/S0165168409003466}
\BIBentrySTDinterwordspacing

\bibitem{rrstap}
R.~Fa and R.~C. De~Lamare, ``Reduced-rank stap algorithms using joint iterative
  optimization of filters,'' \emph{IEEE Transactions on Aerospace and
  Electronic Systems}, vol.~47, no.~3, pp. 1668--1684, 2011.

\bibitem{wljio}
N.~Song, W.~U. Alokozai, R.~C. de~Lamare, and M.~Haardt, ``Adaptive widely
  linear reduced-rank beamforming based on joint iterative optimization,''
  \emph{IEEE Signal Processing Letters}, vol.~21, no.~3, pp. 265--269, 2014.

\bibitem{okspme}
H.~Ruan and R.~C. de~Lamare, ``Robust adaptive beamforming based on low-rank
  and cross-correlation techniques,'' \emph{IEEE Transactions on Signal
  Processing}, vol.~64, no.~15, pp. 3919--3932, 2016.

\bibitem{mcg}
\BIBentryALTinterwordspacing
L.~Wang, \emph{\BIBforeignlanguage{English}{IET Signal Processing}}, vol.~4,
  pp. 686--697(11), December 2010. [Online]. Available:
  \url{https://digital-library.theiet.org/content/journals/10.1049/iet-spr.2009.0243}
\BIBentrySTDinterwordspacing

\bibitem{eljio}
\BIBentryALTinterwordspacing
R.~de~Lamare, ``\BIBforeignlanguage{English}{Adaptive reduced-rank lcmv
  beamforming algorithms based on joint iterative optimisation of filters},''
  \emph{\BIBforeignlanguage{English}{Electronics Letters}}, vol.~44, pp.
  565--567(2), April 2008. [Online]. Available:
  \url{https://digital-library.theiet.org/content/journals/10.1049/el_20080627}
\BIBentrySTDinterwordspacing

\bibitem{wcccm}
\BIBentryALTinterwordspacing
L.~Landau, R.~C. de~Lamare, and M.~Haardt, ``Robust adaptive beamforming
  algorithms using the constrained constant modulus criterion,'' \emph{IET
  Signal Processing}, vol.~8, no.~5, pp. 447--457, 2014. [Online]. Available:
  \url{https://ietresearch.onlinelibrary.wiley.com/doi/abs/10.1049/iet-spr.2013.0166}
\BIBentrySTDinterwordspacing

\bibitem{ccmjio}
L.~Wang, R.~C. de~Lamare, and M.~Yukawa, ``Adaptive reduced-rank constrained
  constant modulus algorithms based on joint iterative optimization of filters
  for beamforming,'' \emph{IEEE Transactions on Signal Processing}, vol.~58,
  no.~6, pp. 2983--2997, 2010.

\bibitem{ccmavf}
L.~Wang and R.~C. de~Lamare, ``Adaptive constrained constant modulus algorithm
  based on auxiliary vector filtering for beamforming,'' \emph{IEEE
  Transactions on Signal Processing}, vol.~58, no.~10, pp. 5408--5413, 2010.

\bibitem{arh}
F.~G. Almeida~Neto, R.~C. De~Lamare, V.~H. Nascimento, and Y.~V. Zakharov,
  ``Adaptive reweighting homotopy algorithms applied to beamforming,''
  \emph{IEEE Transactions on Aerospace and Electronic Systems}, vol.~51, no.~3,
  pp. 1902--1915, 2015.

\bibitem{kaesprit}
S.~F.~B. Pinto and R.~C. de~Lamare, ``Multistep knowledge-aided iterative
  esprit: Design and analysis,'' \emph{IEEE Transactions on Aerospace and
  Electronic Systems}, vol.~54, no.~5, pp. 2189--2201, 2018.

\bibitem{rdrls}
Y.~Yu, H.~Zhao, R.~C. de~Lamare, Y.~Zakharov, and L.~Lu, ``Robust distributed
  diffusion recursive least squares algorithms with side information for
  adaptive networks,'' \emph{IEEE Transactions on Signal Processing}, vol.~67,
  no.~6, pp. 1566--1581, 2019.

\bibitem{lrcc}
H.~Ruan and R.~C. de~Lamare, ``Distributed robust beamforming based on low-rank
  and cross-correlation techniques: Design and analysis,'' \emph{IEEE
  Transactions on Signal Processing}, vol.~67, no.~24, pp. 6411--6423, 2019.

\bibitem{mestre2006finite}
X.~Mestre and M.~A. Lagunas, ``Finite sample size effect on minimum variance
  beamformers: Optimum diagonal loading factor for large arrays,'' \emph{IEEE
  Transactions on Signal Processing}, vol.~54, no.~1, pp. 69--82, 2006.

\bibitem{kukrer2014generalised}
O.~Kukrer and S.~Mohammadzadeh, ``Generalised loading algorithm for adaptive
  beamforming in ulas,'' \emph{Electronics Letters}, vol.~50, no.~13, pp.
  910--912, 2014.

\bibitem{huang2012modified}
F.~Huang, W.~Sheng, and X.~Ma, ``Modified projection approach for robust
  adaptive array beamforming,'' \emph{Signal Processing}, vol.~92, no.~7, pp.
  1758--1763, 2012.

\bibitem{xie2014fast}
H.~Xie, D.~Feng, and H.~Yu, ``Fast and robust adaptive beamforming method based
  on l1-norm constraint for large array,'' \emph{Electronics Letters}, vol.~51,
  no.~1, pp. 98--99, 2014.

\bibitem{yu2010robust}
Z.~L. Yu, Z.~Gu, J.~Zhou, Y.~Li, W.~Ser, and M.~H. Er, ``A robust adaptive
  beamformer based on worst-case semi-definite programming,'' \emph{IEEE
  Transactions on Signal Processing}, vol.~58, no.~11, pp. 5914--5919, 2010.

\bibitem{nai2011iterative}
S.~E. Nai, W.~Ser, Z.~L. Yu, and H.~Chen, ``Iterative robust minimum variance
  beamforming,'' \emph{IEEE Trans. on Signal Process.}, vol.~59, no.~4, pp.
  1601--1611, 2011.

\bibitem{yang2017modified}
H.~Yang, W.~Li, and D.~Cao, ``A modified robust algorithm against large look
  direction error based on interference-plus-noise covariance matrix
  reconstruction and steering vector double estimation,'' in \emph{Progress in
  Electromagnetics Research Symposium-Fall (PIERS-FALL), 2017}.\hskip 1em plus
  0.5em minus 0.4em\relax IEEE, 2017, pp. 615--620.

\bibitem{gu2012robust}
Y.~Gu and A.~Leshem, ``Robust adaptive beamforming based on interference
  covariance matrix reconstruction and steering vector estimation,'' \emph{IEEE
  Transactions on Signal Processing}, vol.~60, no.~7, pp. 3881--3885, 2012.

\bibitem{yuan2017robust}
X.~Yuan and L.~Gan, ``Robust adaptive beamforming via a novel subspace method
  for interference covariance matrix reconstruction,'' \emph{Signal
  Processing}, vol. 130, pp. 233--242, 2017.

\bibitem{mohammadzadeh2018modified}
S.~Mohammadzadeh and O.~Kukrer, ``Modified robust capon beamforming with
  approximate orthogonal projection onto the signal-plus-interference
  subspace,'' \emph{Circuits, Systems, and Signal Processing}, vol.~37, no.~12,
  pp. 5351--5368, 2018.

\bibitem{ruan2014robust}
H.~Ruan and R.~C. de~Lamare, ``Robust adaptive beamforming using a
  low-complexity shrinkage-based mismatch estimation algorithm.'' \emph{IEEE
  Signal Process. Lett.}, vol.~21, no.~1, pp. 60--64, 2014.

\bibitem{gu2014robust}
Y.~Gu, N.~A. Goodman, S.~Hong, and Y.~Li, ``Robust adaptive beamforming based
  on interference covariance matrix sparse reconstruction,'' \emph{Signal
  Processing}, vol.~96, pp. 375--381, 2014.

\bibitem{huang2015robust}
L.~Huang, J.~Zhang, X.~Xu, and Z.~Ye, ``Robust adaptive beamforming with a
  novel interference-plus-noise covariance matrix reconstruction method.''
  \emph{IEEE Trans. Signal Process.}, vol.~63, no.~7, pp. 1643--1650, 2015.

\bibitem{chen2018robust}
P.~Chen, Y.~Yang, Y.~Wang, and Y.~Ma, ``Robust adaptive beamforming with sensor
  position errors using weighted subspace fitting-based covariance matrix
  reconstruction,'' \emph{Sensors}, vol.~18, no.~5, p. 1476, 2018.

\bibitem{chen2015robust}
F.~Chen, F.~Shen, and J.~Song, ``Robust adaptive beamforming using
  low-complexity correlation coefficient calculation algorithms,''
  \emph{Electronics Letters}, vol.~51, no.~6, pp. 443--445, 2015.

\bibitem{zhang2016interference}
Z.~Zhang, W.~Liu, W.~Leng, A.~Wang, and H.~Shi, ``Interference-plus-noise
  covariance matrix reconstruction via spatial power spectrum sampling for
  robust adaptive beamforming,'' \emph{IEEE Signal Processing Letters},
  vol.~23, no.~1, pp. 121--125, 2016.

\bibitem{mohammadzadeh2018adaptive}
S.~Mohammadzadeh and O.~Kukrer, ``Adaptive beamforming based on theoretical
  interference-plus-noise covariance and direction-of-arrival estimation,''
  \emph{IET Signal Processing}, 2018.

\bibitem{zheng2018covariance}
Z.~Zheng, Y.~Zheng, W.-Q. Wang, and H.~Zhang, ``Covariance matrix
  reconstruction with interference steering vector and power estimation for
  robust adaptive beamforming,'' \emph{IEEE Transactions on Vehicular
  Technology}, vol.~67, no.~9, pp. 8495--8503, 2018.

\bibitem{li2003robust}
J.~Li, P.~Stoica, and Z.~Wang, ``On robust capon beamforming and diagonal
  loading,'' \emph{IEEE Trans. on Signal Process.}, vol.~51, no.~7, pp.
  1702--1715, 2003.

\bibitem{somasundaram2014degradation}
S.~D. Somasundaram and A.~Jakobsson, ``Degradation of covariance
  reconstruction-based robust adaptive beamformers,'' in \emph{Sensor Signal
  Process. for Defence (SSPD), 2014}.\hskip 1em plus 0.5em minus 0.4em\relax
  IEEE, 2014, pp. 1--5.

\bibitem{wang2016robust}
Y.~Wang, Q.~Bao, and Z.~Chen, ``Robust adaptive beamforming using iaa-based
  interference-plus-noise covariance matrix reconstruction,'' \emph{Electronics
  Letters}, vol.~52, no.~13, pp. 1185--1186, 2016.

\bibitem{mohammadzadeh2020maximum}
S.~Mohammadzadeh, V.~H. Nascimento, R.~C. de~Lamare, and O.~Kukrer, ``Maximum
  entropy-based interference-plus-noise covariance matrix reconstruction for
  robust adaptive beamforming,'' \emph{IEEE Signal Processing Letters},
  vol.~27, pp. 845--849, 2020.

\bibitem{sun2021robust}
S.~Sun and Z.~Ye, ``Robust adaptive beamforming based on a method for steering
  vector estimation and interference covariance matrix reconstruction,''
  \emph{Signal Processing}, vol. 182, p. 107939, 2021.

\bibitem{zhu2020robust}
X.~Zhu, X.~Xu, and Z.~Ye, ``Robust adaptive beamforming via subspace for
  interference covariance matrix reconstruction,'' \emph{Signal Processing},
  vol. 167, p. 107289, 2020.

\bibitem{zhang2020rcb}
P.~Zhang, Z.~Yang, G.~Liao, G.~Jing, and T.~Ma, ``A rcb-like steering vector
  estimation method based on interference matrix reduction,'' \emph{IEEE
  Transactions on Aerospace and Electronic Systems}, 2020.

\bibitem{capon1969high}
J.~Capon, ``High resolution frequency wavenumber spectrum analysis,''
  \emph{Proceedings of the IEEE}, vol.~57, no.~8, pp. 1408--1418, 1969.

\bibitem{mohammadzadeh2019robust}
S.~Mohammadzadeh and O.~Kukrer, ``Robust adaptive beamforming based on
  covariance matrix and new steering vector estimation,'' \emph{Signal, Image
  and Video Processing}, vol.~13, no.~5, pp. 853--860, 2019.

\bibitem{somasundaram2011evaluation}
S.~D. Somasundaram and N.~H. Parsons, ``Evaluation of robust capon beamforming
  for passive sonar,'' \emph{IEEE Journal of Oceanic Engineering}, vol.~36,
  no.~4, pp. 686--695, 2011.

\bibitem{stoica2005spectral}
P.~Stoica, R.~L. Moses \emph{et~al.}, \emph{Spectral Analysis of
  Signals}.\hskip 1em plus 0.5em minus 0.4em\relax Pearson Prentice Hall Upper
  Saddle River, NJ, 2005.

\bibitem{gu2019adaptive}
Y.~Gu and Y.~D. Zhang, ``Adaptive beamforming based on interference covariance
  matrix estimation,'' in \emph{2019 53rd Asilomar Conference on Signals,
  Systems, and Computers}.\hskip 1em plus 0.5em minus 0.4em\relax IEEE, 2019,
  pp. 619--623.

\bibitem{grant2008cvx}
M.~Grant, S.~Boyd, and Y.~Ye, ``Cvx: Matlab software for disciplined convex
  programming,'' 2008.

\bibitem{besson2000decoupled}
O.~Besson and P.~Stoica, ``Decoupled estimation of doa and angular spread for a
  spatially distributed source,'' \emph{IEEE Transactions on Signal
  Processing}, vol.~48, no.~7, pp. 1872--1882, 2000.

\bibitem{gershman1999robust}
A.~B. Gershman, ``Robust adaptive beamforming in sensor arrays,'' \emph{Int. J.
  Electron. Commun.}, vol.~53, pp. 305--314, 1999.

\end{thebibliography}

\end{document}